\pdfoutput=1
\documentclass[amssymb,amsmath,prl,twocolumn,superscriptaddress,aps,floatfix]{revtex4-2}

\usepackage[caption=false]{subfig}
\usepackage{graphicx}
\usepackage{bm}
\usepackage{amssymb,amsmath}
\usepackage{mathrsfs}
\usepackage{latexsym}
\usepackage{color}
\usepackage[normalem]{ulem}
\usepackage{dcolumn}
\usepackage[colorlinks,urlcolor=dodgerblue,citecolor=dodgerblue,linkcolor=dodgerblue,pdfusetitle]{hyperref}
\usepackage[usenames,dvipsnames]{xcolor}
\usepackage[utf8]{inputenc}
\usepackage{microtype}
\usepackage{etoolbox}
\usepackage{accents}
\usepackage{multirow}
\usepackage{orcidlink}

\allowdisplaybreaks

\definecolor{ZurichBlue}{rgb}{.255,.41,.884} 		% RoyalBlue of svgnames
\definecolor{ZurichRed}{rgb}{0.9, 0.1, 0} 			% Red of svgnames
\definecolor{ZurichGreen}{rgb}{.196,.504,.396} 		% LimeGreen of svgnames
\definecolor{ZurichYellow}{rgb}{1,.648,0} 			% Orange of svgnames
\definecolor{dodgerblue}{rgb}{0.12, 0.56, 1.0}
\definecolor{azure}{rgb}{0.0, 0.5, 1.0}
\definecolor{alizarincrimson}{rgb}{0.82, 0.1, 0.26}
\definecolor{mediumpurple}{rgb}{0.58, 0.44, 0.86}
\definecolor{lasallegreen}{rgb}{0.03, 0.47, 0.19}
\definecolor{my_gray}{rgb}{0,0,0}
\definecolor{uniwienblue}{HTML}{006699}
\definecolor{walesred}{HTML}{ff0038}
\definecolor{myorange}{HTML}{FF6C0C}
\definecolor{ferngreen}{HTML}{009246}
\definecolor{scarletred}{HTML}{CD212A}
\definecolor{grblue}{HTML}{1269C7}

\definecolor{dodgerblue}{HTML}{1E90FF}
\definecolor{viennared}{HTML}{DA0A14}
\definecolor{ctorange}{HTML}{FF6C0C}
\definecolor{wales}{HTML}{ff0038}
\definecolor{benettongreen}{HTML}{009421}
\definecolor{ferrarired}{HTML}{ff2800}
\definecolor{austriawienpurple}{HTML}{441678}
\definecolor{gray}{HTML}{F0F0F0}
\definecolor{LightCyan}{rgb}{0.88,1,1}
\newcolumntype{a}{>{\columncolor{gray}}c}
\newcolumntype{b}{>{\columncolor{white}}c}

\hypersetup{
     hypertexnames=false,
     colorlinks=true,
     linkcolor=dodgerblue,
     filecolor=dodgerblue,
     citecolor = dodgerblue,      
     urlcolor=dodgerblue,
     }
     
\ifdefined\cleanrevision
\colorlet{red}{black}
\colorlet{blue}{black}
\colorlet{magenta}{black}
\colorlet{dodgerblue}{black}
\colorlet{ferrarired}{black}
\colorlet{lasallegreen}{black}
\fi

\ifdefined\cleanrevision
\newcommand{\green}[1]{#1}
\else
\newcommand{\green}[1]{\textcolor{lasallegreen}{#1}}
\fi

\newcommand{\citeme}[1]{\textcolor{red}{[CITEME!]}}

\newcommand{\totalangle}{\chi}

\newcommand{\lhat}{\hat{\bm{L}}}
\newcommand{\pinhat}{\hat{\bm{p}}_{\rm in}}
\newcommand{\bhat}{\hat{\bm{b}}}

\makeatletter
\def\maketitle{\@author@finish \title@column\titleblock@produce \suppressfloats[t]}
\makeatother

\newcommand{\Bham}{\affiliation{School of Physics and Astronomy and Institute for Gravitational Wave Astronomy, University of Birmingham, Edgbaston, Birmingham, B15 2TT, United Kingdom}}

\begin{document}

\title{Binary black hole scattering with generic spins}

%---------- Authors
\author{Adam Clark 
\orcidlink{0009-0001-0551-2481}}
\email{axc157@student.bham.ac.uk}
\Bham

\author{Geraint Pratten 
\orcidlink{0000-0003-4984-0775}}
\email{G.Pratten@bham.ac.uk}
\Bham

\author{Patricia Schmidt 
\orcidlink{0000-0003-1542-1791}}
\email{P.Schmidt@bham.ac.uk}
\Bham

\date{\today}

%---------- Abstract
\begin{abstract}
    We study generic-spin black-hole scattering with orbital-plane precession, whose dynamics require more than one scattering angle.
    From numerical relativity (NR), we extract the polar scattering angle and precession of the orbital plane, the latter characterised by asymptotic Euler angles.
    A dictionary relating post-Minkowskian (PM) observables to scattering and Euler angles is presented, enabling the first precessing PM-NR comparisons.
    Our results provide a key benchmark for generic-spin resummations and provide the foundations for exploring boundary-to-bound maps in the generic case.
\end{abstract}

\maketitle

%---------- Introduction
\noindent{\bf \em Introduction}. 
Gravitational waves (GWs) from compact binary coalescences provide a unique probe of strong-field gravitational dynamics and the nature of the interacting bodies.
Of all their properties, spin is one of the most fundamental~\cite{Carter:1971zc}, encoding valuable information on the astrophysical origin of black holes and neutron stars across the Universe~\cite{Vitale:2020aaz}.
Spin adds significant complexity to the emitted radiation as, for generic spin orientations, spin-orbit and spin-spin couplings drive the precession of the spins and the orbital plane, breaking planar symmetry~\cite{Lense:1918zz, Barker:1975ae, Apostolatos:1994mx, Kidder:1995zr}.
Moreover, recent observations show evidence for eccentricity~\cite{Morras:2025xfu, Planas:2025plq, Gupte:2024jfe, Kacanja:2025kpr, Gamba:2021gap} adding a significant further modelling challenge. 
In eccentric binaries, the orbital velocity peaks at periastron, precisely where the small-velocity post-Newtonian (PN) expansion~\cite{Blanchet:2013haa} is least reliable.
Modelling the interplay between eccentricity and precession in the strong field will be essential in interpreting observations with current~\cite{KAGRA:2021vkt, LIGOScientific:2025slb} and next-generation GW detectors~\cite{ET:2025xjr, Evans:2021gyd}.

Post-Minkowskian (PM) theory offers a complementary route as a weak-field expansion in $G$ with no restriction on velocity~\cite{Khalil:2022ylj}. Each order retains full velocity dependence, resumming an infinite tower of PN terms~\cite{Bertotti:1956pxu, Westpfahl:1979aa, Damour:2016gwp, Damour:2017zjx, Bern:2019nnu}.
Recent progress exploiting the classical limit of quantum gravitational scattering has driven PM calculations to remarkably high orders~\cite{Damour:2017zjx, Bern:2019nnu, Driesse:2024feo, Driesse:2026qiz}. 
A range of theoretical methods have been developed, including scattering amplitudes~\cite{Bern:2019crd, Bern:2021dqo}, worldline effective field theory~\cite{Kalin:2020mvi, Dlapa:2021npj}, and worldline quantum field theory~\cite{Mogull:2020sak, Driesse:2024feo}.
These approaches give direct access to gauge-invariant observables characterising the two-body dynamics, such as the scattering angle~\cite{Damour:2016gwp} and the radial action~\cite{Bern:2021dqo}. 
These quantities feed directly into effective-one-body (EOB) models for bound and unbound orbits~\cite{Buonanno:1998gg, Damour:2016gwp, Buonanno:2024byg, Buonanno:2024vkx, Damour:2025uka}, allowing us to transcribe high-order PM information into state-of-the-art models.

\begin{figure}[tbp]
    \centering
    \includegraphics[width=0.47 \textwidth]{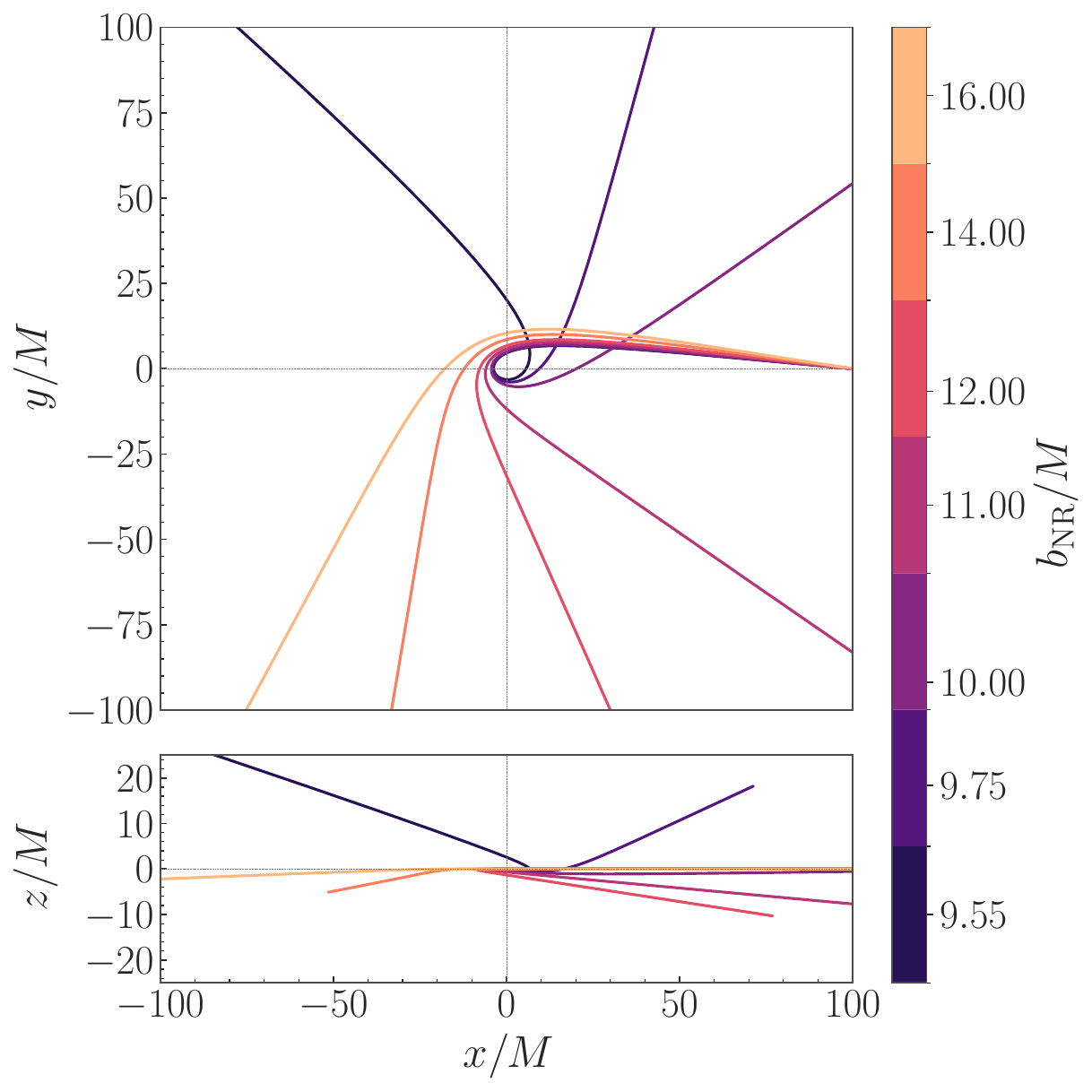}
    \caption{Relative coordinate trajectories of equal-mass black-hole scattering simulations for different impact parameters. The ingoing motion lies in the $xy$-plane (top), while spin-orbit misalignment generates out-of-plane motion (bottom).}
    \label{fig:impact-trajectory-plot}
\end{figure}

PM scattering calculations typically provide the covariant momentum impulse $\Delta p^\mu$ and the spin kick $\Delta S^\mu$. 
Here, we complete that set with explicit expressions for the angular momentum kick $\Delta L^\mu$, using known results for $\Delta p^\mu$ and $\Delta S^\mu$ to $\mathcal{O}(G^3S^2)$~\cite{Jakobsen:2022zsx, Akpinar:2024meg} and $\mathcal{O}(G^4S)$~\cite{Jakobsen:2023hig}. 
Together, these variables encode the transition between asymptotic states~\cite{Bern:2020buy,Jakobsen:2021zvh,Jakobsen:2022zsx,Akpinar:2025bkt}. 
For aligned spins, the process is planar and the scattering angle determines the conservative dynamics up to canonical transformations~\cite{Damour:2017zjx,Kosower:2018adc,Cheung:2018wkq,Mogull:2020sak,Damour:2022ybd,Bern:2019nnu}. 
For generic spins, $\Delta L^\mu$ encodes the precession of the orbital plane and the scattering angle inherits a non-planar, polar component.
Numerical relativity (NR) simulations, on the other hand, provide a non-perturbative benchmark for these PM predictions, allowing us to explore the breakdown of the expansion.
To date, however, PM--NR comparisons have only focused on the nonspinning and aligned-spin sector~\cite{Damour:2014afa,Rettegno:2023ghr,Hopper:2022rwo,Swain:2024ngs,Albanesi:2024xus,Long:2025nmj,Long:2025tvk}. 
A generalisation to the full non-planar dynamics, as shown in Fig.~\ref{fig:impact-trajectory-plot}, requires a common asymptotic frame and spin convention together with a common set of observables that distinguish momentum deflection from the precession of the orbital plane.

In this Letter, we construct this dictionary and derive explicit finite-mass-ratio expressions for the polar scattering angle in both the \textit{canonical} and \textit{covariant} spin supplementary conditions (SSC). 
We relate the scattering angles to the asymptotic orbital-plane Euler angles that are commonly used in waveform models~\cite{Schmidt:2010it,Schmidt:2012rh,Pratten:2020ceb}, and perform the first direct PM--NR comparison of the non-planar sector. 
Our series of numerical simulations reveals a polar scattering angle that grows to a maximum before turning over and reversing sign, while the orbital-plane tilt grows monotonically. 
The PM-expanded scattering angles fail to capture this behaviour. 
However, we do find a similar structure in the scattering of Kerr geodesics, and our NR simulations establish that this behaviour persists in the comparable-mass regime. 
The combined angular observables provide a benchmark for generic-spin resummations and a basis for extending boundary-to-bound maps~\cite{Kalin:2019rwq} to precessing dynamics.

We use geometric units, $G=c=1$, where the black hole masses are denoted by $m_i$, with $i=1,2$, the total mass by $M=m_1 + m_2$, and the spin angular momentum by ${\bm S}_i = m_i {\bm a}_i = m_i^2 {\bm \chi}_i$, where ${\bm a}_i$ is the Kerr spin and $\bm{\chi}_i$ is the dimensionless spin. Throughout we use the metric signature $(+, -, -, -)$. \\

\begin{figure}[tbp]
    \centering
    \includegraphics[width=0.92\columnwidth]{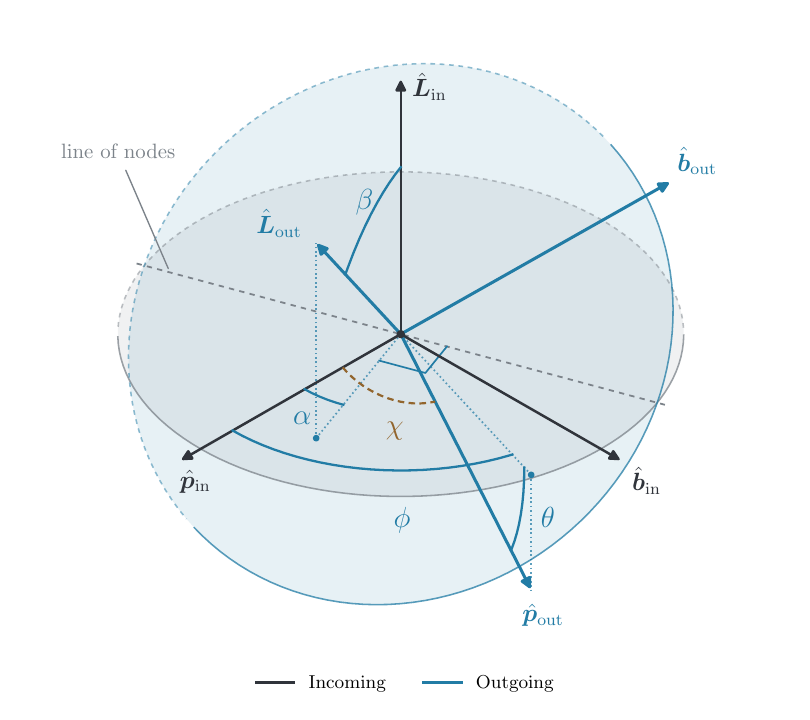}
    \caption{
    Asymptotic scattering geometry for generic spins. We show the incoming (grey) and outgoing (blue) orbital planes. The dotted vertical lines project $\bm p_{\rm out}$ and $\bm L_{\rm out}$ onto the incoming plane, defining $\phi$ and $\alpha$. 
    The orbital-plane tilt $\beta$ is the angle between $\bm L_{\rm in}$ and $\bm L_{\rm out}$. The line-of-nodes (grey dashed) lies at an azimuth of $\alpha\pm\pi/2$. The total scattering angle is defined by $\cos\chi = \hat{\bm{p}}_{\rm in} \cdot \hat{\bm{p}}_{\rm out}$. 
    }
    \label{fig:scattering-geometry}
\end{figure}

% ---------- Geometric Setup
\noindent{\bf \em Geometry of Scattering and Precession}. 
For spins (anti-)aligned with the orbital angular momentum $\bm L$, black-hole scattering is planar in the centre-of-mass (CoM) frame, taken here to be the $xy$-plane. 
Generic spin orientations break this reflection symmetry, with spin-orbit couplings driving the evolution of $\bm L$, and hence the scattering plane, between the ingoing and outgoing states. 
The total angular momentum $\bm J$ remains conserved since radiation reaction is neglected~\cite{Barker:1975ae,Kidder:1995zr,Apostolatos:1994mx}.
The initial and final scattering states are specified at past ($i^-$) and future ($i^+$) timelike infinity~\cite{Compere:2023qoa}. 
At $i^-$, the relative momentum $\bm p_{\rm in}$ and impact parameter $\bm{b}_{\rm in}$ span the initial scattering plane, with its normal
$\bm L_{\rm in}=\bm p_{\rm in}\times\bm b_{\rm in}$. 
At $i^+$, the state is specified by $\bm p_{\rm out}$ and $\bm L_{\rm out}$.
In this Letter, we use a right-handed triad, $\{\pinhat,\bhat_{\rm in},\hat{\bm L}_{\rm in}\}$.
The asymptotic geometry is summarised in Fig.~\ref{fig:scattering-geometry}.

The scattering dynamics for generic spins can be described by three non-independent angles $(\chi, \phi, \theta)$.
The azimuthal and polar scattering angles $(\phi,\theta)$ give the momentum deflection in spherical coordinates for the initial CoM frame, parametrising the dynamics.
In our triad, the outgoing momentum can be parametrised as
\begin{align}
 \hat{\bm p}_{\rm out} &= \cos\theta(\cos\phi\,\pinhat+\sin\phi\,\bhat_{\rm in}) -\sin\theta\,\hat{\bm L}_{\rm in} .
\label{eq:momentum-decomposition}
\end{align}
The \emph{total} scattering angle, $\cos\chi = \hat{\bm{p}}_{\rm in} \cdot \hat{\bm{p}}_{\rm out}$, generalises the aligned-spin scattering angle, with the relation $\cos\chi=\cos\theta\cos\phi$ determining $\chi$, with $\chi=\phi$ in the aligned-spin limit.

In bound systems, orbital precession is commonly described by three Euler angles
~\cite{Apostolatos:1994mx, Buonanno:2002fy, Arun:2008kb, Schmidt:2010it, OShaughnessy:2011pmr, Boyle:2011gg}.
For scattering, we instead introduce two \textit{asymptotic} Euler angles, $(\alpha, \beta)$, which encode the change of $\bm{L}$ from $i^-$ to $i^+$:
\begin{equation} \label{eq:euler-angle-definition}
 \begin{aligned}
    \alpha &= \arctan\!\left(\frac{\hat{\bm L}_{\rm out}\cdot\bhat_{\rm in}}
    {\hat{\bm L}_{\rm out}\cdot\pinhat}\right),\\
    \beta &= \arccos\!\left(\hat{\bm L}_{\rm out}\cdot\hat{\bm L}_{\rm in}\right).
 \end{aligned}
\end{equation}
The quadrant of $\alpha$ is fixed by the relative signs of $\hat{\bm L}_{\rm out}\cdot\pinhat$ and $\hat{\bm L}_{\rm out}\cdot\bhat_{\rm in}$.
Unlike the bound case, no third asymptotic Euler angle is needed since the residual rotation about $\bm L_{\rm out}$ is fixed by $\bm p_{\rm out}$.
Since $\bm p_{\rm out}$ and $\bm L_{\rm out}$ are orthogonal, the scattering and precession angles satisfy the following geometric relation:
\begin{align} \label{eq:theta-beta-relation}
    \tan{\theta} = \tan{\beta} \cos{(\phi - \alpha)}.
\end{align}
This relation connects the out-of-plane momentum deflection with the in-plane rotation of $\bm p_{\rm out}$. 
In particular, $\theta=0$ does not imply $\beta=0$, as the orbital plane can still precess throughout the scattering process, even when there is no out-of-plane impulse.
For $0<\beta<\pi/2$, $\theta$ vanishes whenever $\phi - \alpha = (2n+1)\pi/2$, with $n \in \mathbb{Z}$.
In this range, $\theta$ changes sign as $\phi-\alpha$ crosses an odd multiple of $\pi/2$ and can vary non-monotonically along a scattering sequence. The bound $|\theta| \le \min(\beta, \pi - \beta)$ holds for all $0\le\beta\le\pi$. \\

%---------- PM Scattering
\noindent{\bf \em Post-Minkowskian Scattering}. The state-of-the-art nonspinning results are the complete conservative $5$PM ($\mathcal{O}(G^5)$) \green{\cite{Driesse:2024xad, Bern:2025wyd, Driesse:2026qiz}} and the radiative sector to $\mathcal{O}(G^5 \nu^1)$ \green{\cite{Driesse:2024feo}}, where $\nu=m_1m_2/M^2$ is the symmetric mass ratio counting self-force order.
In the spinning sector, results exist at a range of loop orders, in particular $\mathcal{O}(G^4 S^1)$ \cite{Jakobsen:2023hig}, $\mathcal{O}(G^3 S^4)$ \cite{Akpinar:2025bkt, Haddad:2025cmw}, conjectured higher spin results at $\mathcal{O}(G^2)$ \cite{Bohnenblust:2024hkw, Aoude:2022thd, Aoude:2022trd, Bautista:2023szu}, and all-orders-in-spin at $\mathcal{O}(G^1)$ \cite{Vines:2017hyw}.
The BHs have four-momenta $p_i^\mu$, the CoM energy is defined by $E^2 = s = (p_1 + p_2)^2$ and we define the total and relative momenta as $E\hat{P}^\mu = p_1^\mu + p_2^\mu$, $p_{\rm rel}^\mu = (1/ E) (E_2 p_1^\mu - E_1 p_2^\mu)$, where $E_i = \hat{P}^\mu p_{i \mu}$.
We define $\Gamma=E/M$, and the relative Lorentz factor $\gamma=(s-m_1^2-m_2^2)/(2m_1m_2)$.
Hereafter we work in the CoM frame, defined by $\hat{P}^\mu = (1, 0, 0, 0)$, where $p_1^\mu = (E_1, \bm{p})$ and $p_2^\mu = (E_2, -\bm{p})$.
The impact parameter $b^\mu = b_2^\mu - b_1^\mu = (0,\bm{b}_{\rm in})$ is
the initial orthogonal separation between the worldlines, with the initial
angular-momentum four-vector being defined as $L^\mu_{\rm in} = (1/E)\epsilon^{\mu}{}_{\nu\rho\sigma} b^\nu p_1^\rho p_2^\sigma = (0,\bm{p}_{\rm in}\times\bm{b}_{\rm in})$.
The initial relative momentum has magnitude $|\bm{p}_{\rm in}| = p_{\infty} = m_1 m_2 \sqrt{\gamma^2 - 1}/E$.
We transform the impulse from the covariant SSC~\cite{Tulczyjew:1959}, commonly used in PM calculations, to the canonical SSC, whose spin variables are canonical~\cite{Pryce:1935ibt, Pryce:1948pf, Newton:1949cq, Vines:2016unv} and naturally suited for comparisons with NR. 

For conservative motion, the total scattering angle $\totalangle$ can be computed from the SSC-shifted impulse~\cite{Jakobsen:2022zsx}, which reduces to
\begin{align}
    \sin{\frac{\totalangle}{2}} = \frac{| \Delta p |}{2 p_{\infty}}.
    \label{eq:chi-angle}
\end{align}
To cleanly isolate the precession sector and facilitate comparisons to NR, we decompose the impulse into the following azimuthal and polar components:
\begin{align}
    \sin\theta = \frac{\hat{L}_{\rm in, \mu}\Delta p^\mu}{p_\infty},\qquad
    \cos\theta\sin\phi=-\frac{\hat{ b}_{\rm in, \mu} \Delta p^\mu}{p_\infty}.
\end{align}
For generic spins, $\phi$ and $\chi$ agree through linear order in the spin, since $\cos\chi=\cos\theta\cos\phi$ and $\theta=\mathcal{O}(S)$. 
The polar scattering angle $\theta$ trivially vanishes for aligned spins. 
For generic orientations, it is obtained by projecting the impulse onto $\hat{ L}_{\rm in}^\mu$, and had previously been derived only in the probe limit~\cite{Gonzo:2023goe}.
Here we derive $\theta$ in the canonical SSC to $\mathcal{O}(G^3 S^2)$ including spin-spin effects~\cite{Jakobsen:2022zsx} and to $\mathcal{O}(G^4 S)$ from the $4$PM results of~\cite{Jakobsen:2023hig}.
The expansion through quadratic order in the spins takes the form
\begin{widetext}
    \begin{equation}
    \begin{split}
        \frac{\theta}{\Gamma} &= \sum_{i=1}^{3} \left( \frac{G M}{|b|} \right)^i \Big[ \Big\{ \frac{a_{1,\mu}}{|b|} \left( c^{(i)}_{1,b} \hat{b}^\mu + c^{(i)}_{1,p} \hat{p}^\mu \right)
              + \frac{a_1^\mu a_1^\nu}{|b|^2} \hat{L}_\mu \left( c^{(i)}_{(11),Lb} \hat{b}_\nu + c^{(i)}_{(11),Lp} \hat{p}_\nu \right)
            + c^{(i)}_{(12),Lp} \frac{a_1^\mu a_2^\nu}{|b|^2} \hat{L}_\mu \hat{p}_\nu + \left( 1 \leftrightarrow 2 \right) \Big\}  \\
        & \hspace{10em} + \frac{\hat{L}_\mu \hat{b}_\nu}{|b|^2} \left( c^{(i)}_{\{12\},Lb} \left( a_1^\mu a_2^\nu + a_2^\mu a_1^\nu \right) + c^{(i)}_{[12],Lb} \left( a_1^\mu a_2^\nu - a_2^\mu a_1^\nu \right) \right) \Big]
        \label{eq:theta-expansion}
    \end{split}
\end{equation}
where $p_{\rm rel}^\mu=(0,\bm p_{\rm in})$ and the coefficients $c_X^{(n)}$ depend on the masses and the Lorentz factor $\gamma$. We provide the coefficients through 3PM in the Supplemental Material and through 4PM in the ancillary files~\cite{Clark:2026GHR}.
\end{widetext}

The asymptotic Euler angles $(\alpha, \beta)$ follow from the change in the orbital angular momentum, $\Delta L^\mu= L_{\rm out}^\mu- L_{\rm in}^\mu$. 
In the canonical SSC, the total angular momentum \emph{vector} splits into orbital and spin parts, $J^\mu = L^\mu + S_1^\mu + S_2^\mu$.
Restricting to conservative dynamics, conservation of $J^\mu$ (i.e., $\Delta J^\mu = 0$) leads directly to
\begin{align}
    \Delta L^{\mu} &= - \left(\Delta S_{1,{\rm can}}^\mu + \Delta S_{2,{\rm can}}^\mu\right).
    \label{eq:DeltaL-spin-kick}
\end{align}
As the canonical spin differs from the covariant spin by a momentum-dependent shift, $\Delta S_{i,{\rm can}}$, and hence $\Delta L^{\mu}$, will depend on the impulse as well as the spin kick.
In our choice of SSC, the angular momentum kick is purely spatial and we now work with three-vectors for clarity.

To extract the Euler angles, we decompose $\bm L_{\rm out}$ following Eq.~\eqref{eq:euler-angle-definition},
\begin{align}
    \Delta\bm L &= |\bm L_{\rm out}|\sin\beta
        (\cos\alpha\,\pinhat+\sin\alpha\,\bhat_{\rm in})\notag\\
        &\qquad+(|\bm L_{\rm out}|\cos\beta-|\bm L_{\rm in}|)\hat{\bm L}_{\rm in} .
\end{align}
The in-plane components fix the precession angle,
\begin{align}
    \tan\alpha = \frac{\bhat_{\rm in}\cdot\Delta\bm L}{\pinhat\cdot\Delta\bm L}.
    \label{eq:alpha-from-spin-kick}
\end{align}
Since conservation of $\bm J$ does not imply
conservation of $|\bm L|$, the orbital-plane tilt is obtained from 
\predisplaypenalty=0
\begin{align}
    \cos\beta = \frac{|\bm L_{\rm in}|+\hat{\bm L}_{\rm in}\cdot\Delta\bm L}
    {|\bm L_{\rm in}+\Delta\bm L|}.
    \label{eq:beta-from-spin-kick}
\end{align}
Together, Eqns.~\eqref{eq:alpha-from-spin-kick} and~\eqref{eq:beta-from-spin-kick} allow us to connect the canonical spin kick to the Euler angles that are typically used in waveform models~\cite{Schmidt:2010it}.

For the remainder of this work, we restrict to initial configurations where the spins are confined to the initial orbital place. 
Reflection through the orbital plane simultaneously reverses the in-plane spin components and the polar deflection, leaving $\phi$ and $\cos\beta$ unchanged, but taking $\theta \rightarrow -\theta$. 
As such, $\theta$ only contains odd powers of spin. All even-in-spin contributions to Eq.~\eqref{eq:theta-expansion} identically vanish. 
Moreover, $c^{(1)}_{1,p}=0$, as the $1$PM impulse only depends on the spin components orthogonal to the incoming velocities~\cite{Vines:2017hyw}. For spin configurations with an out-of-plane component, the even powers of spin enter the polar scattering angle, see the supplemental material for details. \\

\begin{figure*}[t]
    \centering
    \includegraphics[width=0.875\linewidth]{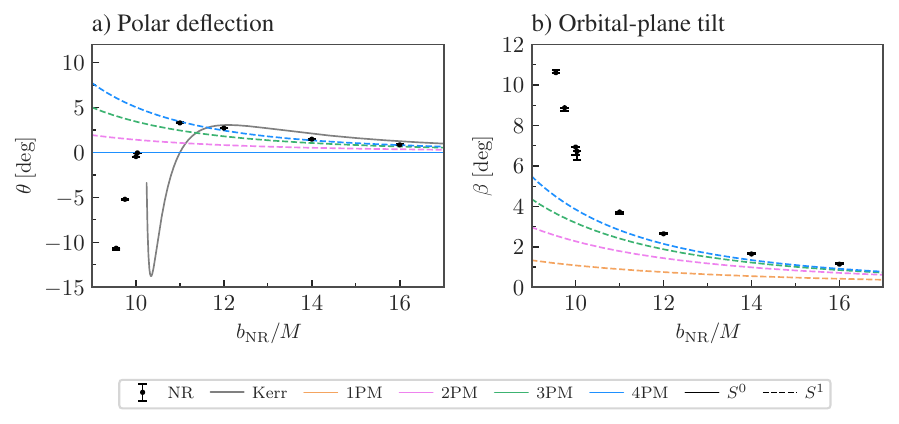}
    \caption{Polar deflection $\theta$ (left) and orbital-plane tilt $\beta$ (right) for the same equal-mass NR sequence, compared with conservative PM predictions through $\mathcal{O}(G^4S)$. The colours denote the PM orders and the line styles denote the spin order. The quadratic-spin contribution to $\theta$ vanishes for these configurations. In grey, we show the Kerr geodesic result for $\theta$. As the impact parameter decreases, the NR polar deflection peaks and reverses sign while the tilt grows. All displayed direct PM-expanded angles remain monotonic. The NR error bars denote the uncertainty due to the spread over extraction polynomial orders.}
    \label{fig:angular-benchmark}
\end{figure*}

%---------- Methodology: NR
\noindent{\bf \em Numerical Relativity Simulations}. We perform equal-mass NR simulations with one spinning and one nonspinning black hole. The NR coordinate frame, $\{ \hat{\bm{x}}, \hat{\bm{y}}, \hat{\bm{z}} \}$, is related to the kinematic triad by $\hat{\bm{x}} = - \hat{\bm{p}}_{\rm in}$, $\hat{\bm{y}} = - \hat{\bm{b}}_{\rm in}$, $\hat{\bm{z}} = \hat{\bm{L}}_{\rm in}$. The spinning black hole has a spin vector $\bm S=0.125\,\hat{\bm x}$ ($|{\bm \chi}_1|=0.5$), which is orthogonal to the initial orbital angular momentum. 
We choose a single-spin series as this isolates the non-planar dynamics without introducing additional spin-spin couplings.
We evolve the spacetime with the moving-punctures method~\cite{Ansorg:2004ds,Campanelli:2005dd,Baker:2005vv}, as implemented in the open-source \textsc{Einstein Toolkit}~\cite{EinsteinToolkit:2025_05} via the \textsc{Cactus} framework~\cite{Cactuscode:web}, using the setup of Ref.~\cite{Rettegno:2023ghr}.

For this sequence, the BHs are initially placed on the $x$-axis at separation $D=100M$, with Arnowitt--Deser--Misner (ADM) linear momenta
\begin{equation}
    \bm{P}_{\rm ADM}^{i} = \pm |\bm{P}_{\rm ADM}|
    \left(-\sqrt{1-\left(\frac{b_{\rm NR}}{D}\right)^2},
    \frac{b_{\rm NR}}{D},0\right),
\end{equation}
where $b_{\rm NR}$ is the impact parameter. We use $|\bm{P}_{\rm ADM}|=0.11456$, giving $\Gamma=1.02281$ and $\gamma=1.09230$, corresponding to an initial relative velocity $v\simeq0.4c$. We relate $|\bm b|$ to $b_{\rm NR}$ by equating the initial angular momenta in the NR and PM descriptions~\cite{Long:2025nmj}, $p_\infty|\bm b|=|\bm P_{\rm ADM}|b_{\rm NR}$. Figure~\ref{fig:impact-trajectory-plot} shows the relative CoM coordinate trajectories for different impact parameters.

We extract the scattering angles by generalising the methods of Refs.~\cite{Damour:2014afa,Rettegno:2023ghr,Swain:2024ngs} to out-of-plane dynamics. 
We parametrise the trajectory using spherical coordinates and extrapolate the incoming and outgoing branches using least-squares polynomial fits stabilised by singular value decomposition.
In the NR frame the azimuthal and polar scattering angles are defined as
\begin{equation}
    \phi=\varphi_{\rm out}-\varphi_{\rm in}-\pi,
    \qquad
    \theta=\Theta_{\rm out}-\Theta_{\rm in},
\end{equation}
where $\varphi_{\rm in/out}$ and $\Theta_{\rm in/out}$ are the asymptotic angles of the ingoing and outgoing trajectories, measured with respect to the NR frame. 
The scattering angles are gauge invariant once the SSC is fixed asymptotically.
A shift in the SSC corresponds to a small rotation of the initial triad in the $\{\bhat_{\rm in}, \lhat_{\rm in}\}$-plane, and thus changes the definition of the scattering angle~\cite{Clark:2026ab}. 
However, for a fixed SSC, coordinate changes that vanish at infinity leave the scattering angles unchanged.

The same fitting procedure can be applied to the components of $\hat{\bm L}_{\rm out}$, yielding the tilt $\beta$ through Eq.~\eqref{eq:euler-angle-definition}.
However, junk radiation prevents a reliable fit for $\bm L_{\rm in}$, so we instead use the initial-data values $\bm L_{\rm in}=(0,0,P_{\rm ADM}b_{\rm NR})$, giving $\beta_{\rm in}=0$. 
Since $\alpha_{\rm in}$ is undefined, we choose the convention $\alpha_{\rm in}=0$. 
Here, the error bars only denote the error introduced by the polynomial order of the fit, and do not include all possible systematics, see Ref.~\cite{Clark:2026ab}.

% ---------- Comparisons
\noindent{\bf \em Comparing NR with analytical predictions}. 
Our NR sequence tests how the polar momentum deflection and orbital-plane tilt evolve as we move from the weak field towards the scatter-capture separatrix, as shown in Fig.~\ref{fig:angular-benchmark}. The PM predictions calculated here are conservative, whereas the NR simulations include radiative corrections. 
In the weak field regime, we find that the $\mathcal{O}(G^4S)$ result captures the magnitude and impact-parameter dependence of the polar deflection up to $b_{\rm NR} \lesssim 12M$.
We note that the $\mathcal{O}(S^2)$ contributions to $\theta$ vanish identically for the configurations here. 
Since $\cos\chi=\cos\theta\cos\phi$, and $\theta=\mathcal{O}(S)$, the total
and azimuthal angles agree within errors throughout, as are shown in
Fig.~\ref{fig:impact-comparison-plot} of the Supplemental Material. 
Our azimuthal angles are largely in agreement with Ref.~\cite{Rettegno:2023ghr}.

As the impact parameter decreases, the NR polar deflection angle reaches a positive maximum near $b_{\rm NR} \simeq 11M$ and then reverses sign. 
The tilt, on the other hand, just increases monotonically, see Fig.~\ref{fig:angular-benchmark}. 
This behaviour also has a simple geometric interpretation through Eq.~\eqref{eq:theta-beta-relation}. 
For $0 < \beta < \pi/2$, the tilt only bounds the polar deflection, $|\theta| \leq \beta$, while its sign and magnitude are determined by the relative azimuth angle $(\phi-\alpha)$. 
Since $\theta$ vanishes whenever the outgoing momentum lies along the line
of nodes, $\phi-\alpha=\pi/2$, the turnover is driven by the azimuthal deflection growing rapidly towards the separatrix. 
As $\phi-\alpha$ passes through $\pi/2$, it leads to a change in sign of $\cos(\phi-\alpha)$, and the polar deflection reverses. 
We note that continued variation of $(\phi - \alpha)$ along the scattering sequence can lead to additional extrema in $\theta$, even if $\beta$ increases monotonically. 
However, none of the direct PM expansions for $\theta$ considered here reproduce the turnover and sign reversal observed in NR. 
The absence of the turnover is a structural property of PM-expanding $\theta$, which reduces to a polynomial in $GM/|b|$ that is only accurate while $(\phi-\alpha)$ is small. 

To explore whether this behaviour also occurs in the exact test-mass dynamics, we consider scattering along Kerr geodesics~\cite{Gonzo:2023goe}. 
We integrate polar Kerr geodesic equations in Mino time~\cite{Mino:2003yg} to predict the azimuthal and polar scattering angles through a single radial libration.  
The geodesic equations are integrated without a PM expansion, retaining their full dependence on $GM/b$ in the test-mass limit. 
We take the incoming momenta to be collinear with the spin axis of the black hole, matching our single-spin configurations. 
In this case, the axial angular momentum vanishes and Eq.~\eqref{eq:theta-beta-relation} reduces to $\tan\theta=\pm\tan\beta\sin\phi$.
The azimuthal and polar scattering angles are shown in the left and right panels of Fig.~\ref{fig:impact-comparison-plot} respectively. 
Although Kerr geodesics overestimate the polar deflection in the weak field, as expected from the $\mathcal{O}(G)$ mass-ratio dependence of Eq.~\eqref{eq:theta-expansion}, they exhibit the same positive maximum and sign seen in NR but also show a negative minimum at smaller values of $b$.
Our closest NR simulation to the separatrix, $b_{\rm NR}=9.5M$, tentatively showed the onset of a second extremum, although its proximity to capture limited the accuracy to which we can confidently extract the scattering angles~\cite{Clark:2026ab}. 
As such we opted to exclude it from the quantitative comparison presented here. 

Successive extrema of $\theta(b)$ relate to the angular phase accumulated during a single radial encounter, though we note that this does not necessarily imply additional radial turning points.
The turnover is therefore a strong-field geometric feature already present
at geodesic order that persists at equal mass.
Since Eq.~\eqref{eq:theta-beta-relation} fixes $\theta$ once $\phi$ and the Euler angles are known, a resummation of the azimuthal deflection that captures its near-separatrix behaviour, combined with the PM Euler angles, should recover the sign reversal. 
The results presented here motivates exploration of resummation strategies that preserve the predicted strong-field behaviour. 
This will be the focus of future work. 

\noindent{\bf \em Discussion.} We have performed the first numerical computation of observables for BH scattering with precessing spins and compared with state-of-the-art analytical results. 
The comparison rests on a dictionary relating the PM impulse to the azimuthal and polar scattering angles and the spin kick to the asymptotic  Euler angles. The latter can be connected to the Euler angles typically  used in bound waveform modelling~\cite{Schmidt:2010it, Pratten:2020ceb}.
A key observation is that the equal-mass NR sequence exhibits a turnover in the polar scattering angle and a sign reversal while the orbital-plane tilt grows monotonically. 
This behaviour is absent from the PM-expanded angles considered here. 
Together, they provide a key strong-field benchmark for generic-spin models. 
An EOB construction combining finite-mass-ratio PM information with Kerr strong-field structure offers a natural route for effectively resumming the PM dynamics~\cite{Damour:2022ybd,Buonanno:2024byg,Rettegno:2023ghr,Clark:2025kvu}.

We further derived a simple, exact, geometric relation connecting the scattering angles with the Euler angles in Eq.~\eqref{eq:theta-beta-relation}, providing valuable insights into the structure of the polar scattering angle: $\theta$ is bounded above by $\beta$ and vanishes whenever the outgoing momentum is collinear with the line of nodes.
Our geometric framework provides the foundations of exploring boundary-to-bound maps on generic orbits~\cite{Kalin:2019rwq,Kalin:2019inp,Cho:2021arx,Gonzo:2023goe,Gonzo:2024xjk}. 
Recent formulations in terms of the radial and on-shell actions~\cite{Akpinar:2025tct,Bern:2021dqo,Haddad:2025cmw,Gonzo:2026yha,Kim:2025gis} provide a unifying framework for such developments.
The results presented here highlight the importance of considering the full degrees of freedom in the generic scattering problem and of understanding the non-perturbative structure of these encounters. 

\label{sec:conclusions}

%----------  Acknowledgments
\noindent{\bf \em Acknowledgments} --  
The authors thank Dogan Akpinar, Christian Chapman-Bird, Lucile Cangemi, Thibault Damour, Riccardo Gonzo, Oliver Long, Piero Rettegno, Shaun Swain and Mao Zeng for useful discussions.
G.P. is supported by a Royal Society University Research Fellowship URF\textbackslash R1\textbackslash 221500 and RF\textbackslash ERE\textbackslash 221015. A.C. is supported by a PhD studentship through these Royal Society grants.
P.S. acknowledges support from a Royal Society Research Grant RG{\textbackslash}R1{\textbackslash}241327.
G.P. and P.S. acknowledge STFC grants ST/V005677/1 and ST/Y00423X/1, and UKSA grant ST/Y004922/1. 
We thank Nordita and the Institute for Advanced Study for their hospitality during ``Amplitudes, Strong-Field Gravity and Resummation'' and ``Prospects in Theoretical Physics,'' respectively, where parts of this work were carried out.
% Computing acknowledgment
Numerical simulations were performed on the University of Birmingham's BlueBEAR HPC, the Sulis Tier 2 HPC at the University of Warwick, funded by EPSRC Grant EP/T022108/1 and the HPC Midlands+ consortium, and the Bondi HPC at the Institute for Gravitational Wave Astronomy.

\vspace{-4mm}

%---------- Bibliography
\bibliography{references}

%---------- END OF MAIN ---------- 

%---------- Supplement
\clearpage
\newpage

\setcounter{equation}{0}
\setcounter{figure}{0}
\renewcommand{\thefigure}{S\arabic{figure}}
\setcounter{table}{0}
\setcounter{page}{1}

\onecolumngrid

\begin{center}
\textbf{Supplemental material for Binary black hole scattering with generic spins}\\[1ex]
Adam Clark, Geraint Pratten, and Patricia Schmidt\\[1ex]
\small School of Physics and Astronomy and Institute for Gravitational Wave Astronomy,\\
University of Birmingham, Edgbaston, Birmingham, B15 2TT, United Kingdom
\end{center}

\section{Azimuthal and total deflection}
\begin{figure}[htb]
    \centering
    \includegraphics[width=0.95\linewidth]{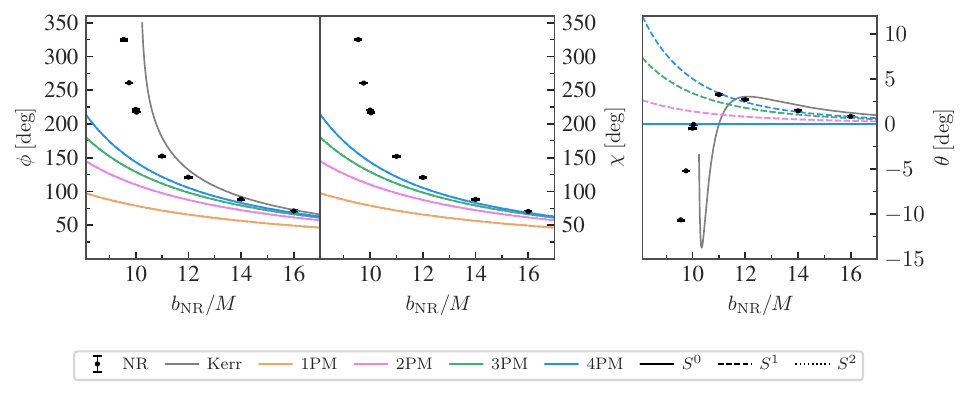}
    \caption{Comparison between NR, PM and the test-mass limit for the series of NR simulations.
    PM data are computed to $\mathcal{O}(G^3 S^2)$ and $\mathcal{O}(G^4 S^1)$. Each spin order is shown at each PM order, with the same colour but different style.
    For $\phi$ and $\chi$ (left), the linear-in-spin terms vanish identically for spins in the initial orbital plane, and the quadratic terms cannot be resolved at this scale. For $\theta$ (right), the quadratic-in-spin terms vanish identically, so the $S^2$ curves coincide with the $S^1$ curves. 
    }
    \label{fig:impact-comparison-plot}
\end{figure}

Figure~\ref{fig:impact-comparison-plot} supplements the polar-deflection and plane-tilt comparison in the main text. It shows the azimuthal and total scattering angles for the same NR sequence, with the polar angle repeated for reference. 

\clearpage
\section{Coordinate motion}

\begin{figure}[htb]
    \centering
    \includegraphics[width=0.5\linewidth]{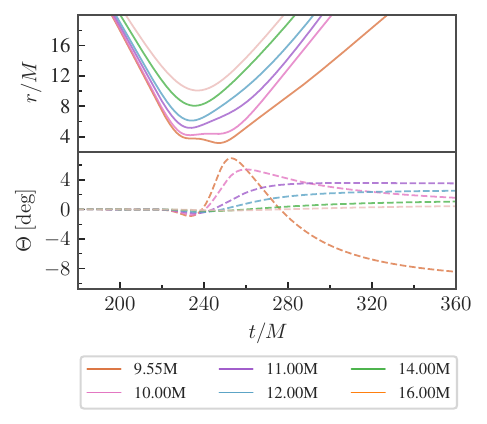}
    \caption{Relative radial (top) and polar (bottom) motion for different $b_{\rm NR}$. The lower panel plots the offset $\Theta(t)-\Theta_{\rm in}$, with $\Theta_{\rm in}=\pi/2$, and labels it as $\Theta$. Its outgoing asymptotic value is the polar scattering angle $\theta$.}
    \label{fig:radial-polar-plot}
\end{figure}

Figure~\ref{fig:radial-polar-plot} shows the relative radial and polar coordinates during the encounter. At smaller impact parameters, the polar motion develops an extremum and can reverse before escape. These finite-separation coordinate features complement the asymptotic polar deflection but are not additional scattering observables. In particular, extrema of the NR coordinate radius should not be identified with roots of the Kerr radial potential.

\clearpage
\section{Coefficients of the polar scattering angle}

In this supplemental material we provide the explicit expressions for the polar scattering angle coefficients in Eq.~\eqref{eq:theta-expansion} to $\mathcal{O}(G^3 S^2)$.
We provide the coefficients in the \emph{canonical} SSC for purely conservative dynamics, which were used to produce the comparisons in the main text. 
We provide a detailed account of the observables in~\cite{Clark:2026ab} and make all of the expressions available at~\cite{Clark:2026GHR}.

Working in the canonical SSC, we define $\gamma_i=E_i/m_i$, $\hat{P}^\mu=(p_1^\mu+p_2^\mu)/E$, and $\mu=m_1m_2/M$, and $|p|  = m_1 m_2 \sqrt{\gamma^2 - 1} / E = \mu \sqrt{\gamma^2 - 1} / \Gamma $, where $M = m_1 + m_2$ is the total mass.
We introduce the mass difference, $\delta = (m_1 - m_2) / M$
Under $(1\leftrightarrow2)$, with $\hat b$, $\hat L$, $\hat p$ held fixed, the coefficients of the particle-2 structures follow from those shown by $\delta\to-\delta$. 
The antisymmetric combination $(\hat L\cdot a_1)(\hat b\cdot a_2)-(\hat L\cdot a_2)(\hat  b\cdot a_1) = -\hat{\boldsymbol p}\cdot(\boldsymbol a_1\times\boldsymbol a_2)$ is odd under the exchange, so its coefficient is odd in $\delta$.

Due to the length of the $4$PM expressions they are omitted here, but are provided in the ancillary files.
The polar scattering angle to quadratic order in the BH spins is then given by the following coefficient expansion:
\begin{equation}
    \begin{split}
        \frac{\theta}{\Gamma} &= \sum_{i=1}^{3} \left( \frac{G M}{|b|} \right)^i \Big[ \Big\{ \frac{a_{1,\mu}}{|b|} \left( c^{(i)}_{1,b} \hat{b}^\mu + c^{(i)}_{1,p} \hat{p}^\mu \right)
            \\ & \qquad + \frac{a_1^\mu a_1^\nu}{|b|^2} \hat{L}_\mu \left( c^{(i)}_{(11),Lb} \hat{b}_\nu + c^{(i)}_{(11),Lp} \hat{p}_\nu \right)
            + c^{(i)}_{(12),Lp} \frac{a_1^\mu a_2^\nu}{|b|^2} \hat{L}_\mu \hat{p}_\nu + \left( 1 \leftrightarrow 2 \right) \Big\}  \\
        & \qquad + \frac{\hat{L}_\mu \hat{b}_\nu}{|b|^2} \left( c^{(i)}_{\{12\},Lb} \left( a_1^\mu a_2^\nu + a_2^\mu a_1^\nu \right) + c^{(i)}_{[12],Lb} \left( a_1^\mu a_2^\nu - a_2^\mu a_1^\nu \right) \right) \Big]
        \label{eq:theta-expansion}
    \end{split}
\end{equation}

% ---------------- 1PM ----------------
\begin{align}
c^{(1)}_{1,b} &= -\frac{2}{(\gamma +1) \sqrt{\gamma ^2-1}}\left[1 + 2 \gamma + \frac{\left(2 \gamma ^2-1\right) (\delta +1)}{\Gamma +1}\right], \qquad
c^{(1)}_{1,p} = 0, \\
c^{(1)}_{(11),Lb} &= \frac{8}{(\gamma +1) \left(\gamma ^2-1\right)} \Bigg[(\gamma -1) (2 \gamma +1) + \frac{-6 \gamma ^2-(\gamma -1) (2 \gamma +1) \delta +\gamma +3}{\Gamma +1}  - \frac{(\gamma -1) \left(2 \gamma ^2-1\right) (\delta +1)}{(\Gamma +1)^2}\Bigg], \\
c^{(1)}_{(11),Lp} &= 0, \qquad
c^{(1)}_{(12),Lp} = 0, \qquad
c^{(1)}_{\{12\},Lb} = -\frac{8 (2 \gamma -1) \Gamma }{\left(\gamma ^2-1\right) (\Gamma +1)}, \qquad
c^{(1)}_{[12],Lb} = 0.
\end{align}

% ---------------- 2PM ----------------
\begin{align}
c^{(2)}_{1,b} &= -\frac{\pi }{8 \left(\gamma ^2-1\right)^{3/2}} \Bigg[-6 - 15 \gamma + 30 \gamma ^2 + 5 \gamma ^3 + \gamma  \left(5 \gamma ^2-3\right) \delta + \frac{6 (\gamma -1) \left(5 \gamma ^2-1\right) (\delta +1)}{\Gamma +1}\Bigg], \\
c^{(2)}_{1,p} &= \frac{1}{\left(\gamma ^2-1\right)^{3/2}}\left[-1 - 4 \gamma + 4 \gamma ^2 + 8 \gamma ^3 + \left(8 \gamma ^3-4 \gamma ^2-4 \gamma +1\right) \delta\right], \\
c^{(2)}_{(11),Lb} &= \frac{3 \pi }{32 \left(\gamma ^2-1\right)^2} \Bigg[49 + 72 \gamma - 342 \gamma ^2 + 200 \gamma ^3 + 5 \gamma ^4 + (\gamma -1) \left(5 \gamma ^3-35 \gamma ^2-17 \gamma +7\right) \delta \nonumber \\
&\qquad + \frac{16 (\gamma -1) \left(-5 \gamma ^3+2 \left(5 \gamma ^2-3\right) \gamma  \nu -45 \gamma ^2-\left(5 \gamma ^3+15 \gamma ^2-9 \gamma -3\right) \delta +9 \gamma +9\right)}{\Gamma +1} \nonumber \\
&\qquad - \frac{48 (\gamma -1)^2 \left(5 \gamma ^2-1\right) (\delta +1)}{(\Gamma +1)^2}\Bigg], \\
c^{(2)}_{(11),Lp} &= \frac{6}{\left(\gamma ^2-1\right)^2} \Bigg[\gamma  \left(4 \gamma ^2-3\right) + (\gamma -1) (2 \gamma -1) (2 \gamma +1) \delta \nonumber \\
&\qquad + \frac{2 (\gamma -1) \left(2 \gamma  \left(2 \gamma ^2-1\right) (\delta +1)-\left(8 \gamma ^3-4 \gamma ^2-4 \gamma +1\right) \nu \right)}{\Gamma +1}\Bigg], \\
c^{(2)}_{(12),Lp} &= \frac{6}{\left(\gamma ^2-1\right)^2} \Bigg[\gamma  \left(4 \gamma ^2-3\right) - (\gamma -1) (2 \gamma -1) (2 \gamma +1) \delta \nonumber \\
&\qquad + \frac{(\gamma -1) \left(2 \left(8 \gamma ^3-4 \gamma ^2-4 \gamma +1\right) \nu +(2 \gamma -1) (2 \gamma +1) (\delta +1)\right)}{\Gamma +1}\Bigg], \\
c^{(2)}_{\{12\},Lb} &= -\frac{3 \pi }{4 \left(\gamma ^2-1\right)^2} \Bigg[6 - 21 \gamma - 18 \gamma ^2 + 35 \gamma ^3 + \frac{2 (\gamma -1) \left(2 \gamma  \left(5 \gamma ^2-3\right) \nu -3 \left(5 \gamma ^2+2 \gamma -1\right)\right)}{\Gamma +1}\Bigg],\\
c^{(2)}_{[12],Lb} &= 0.
\end{align}

% ---------------- 3PM ----------------
\begin{align}
c^{(3)}_{1,b} &= \frac{2}{3 (\gamma +1) \left(\gamma ^2-1\right)^3} \Bigg[-3 \sqrt{\gamma ^2-1} \left(16 \gamma ^5+8 \gamma ^4-24 \gamma ^3-10 \gamma ^2+6 \gamma +1\right) \nonumber \\
&\qquad + \nu\bigg(-2 (\gamma -1) \sqrt{\gamma ^2-1} \left(20 \gamma ^5-94 \gamma ^4-274 \gamma ^3-232 \gamma ^2-112 \gamma -37\right) \nonumber \\
&\qquad + 12 (\gamma +1) (2 \gamma +1) \left(\gamma ^2-1\right) \left(2 \gamma ^3-6 \gamma ^2-3 \gamma -3\right) \operatorname{ArcCosh}(\gamma)\bigg) \nonumber \\
&\qquad + \frac{\delta +1}{\Gamma +1}\bigg(-3 \sqrt{\gamma ^2-1} \left(4 \gamma ^3-4 \gamma -1\right) \left(4 \gamma ^3-4 \gamma +1\right) \nonumber \\
&\qquad + \nu\Big(-2 (\gamma -1) \sqrt{\gamma ^2-1} \left(20 \gamma ^6-28 \gamma ^5-118 \gamma ^4-22 \gamma ^3+98 \gamma ^2+50 \gamma -3\right) \nonumber \\
&\qquad + 12 \left(\gamma ^2-1\right)^2 \left(4 \gamma ^4-12 \gamma ^2-3\right) \operatorname{ArcCosh}(\gamma)\Big)\bigg)\Bigg], \\
c^{(3)}_{1,p} &= \frac{\pi }{4 \left(\gamma ^2-1\right)^{5/2}} \Bigg[3 + 18 \gamma - 27 \gamma ^2 - 80 \gamma ^3 + 30 \gamma ^4 + 70 \gamma ^5 \nonumber \\
&\qquad + \left(70 \gamma ^5-30 \gamma ^4-80 \gamma ^3+27 \gamma ^2+18 \gamma -3\right) \delta - 2 (\gamma -1) \gamma  (2 \gamma +1) \left(5 \gamma ^2-3\right) \nu\Bigg], \\
c^{(3)}_{(11),Lb} &= \frac{2}{15 \left(\gamma ^2-1\right)^4} \Bigg[-5 (\gamma -1) \left(160 \gamma ^7-720 \gamma ^6-504 \gamma ^5+1476 \gamma ^4+600 \gamma ^3-747 \gamma ^2-298 \gamma +45\right) \nonumber \\
&\qquad + 240 (\gamma -1) \left(\gamma ^2-1\right)^{3/2} \left(4 \gamma ^4-12 \gamma ^2-3\right) \operatorname{ArcCosh}(\gamma) \nonumber \\
&\qquad + \delta\bigg(-5 (\gamma -1)^2 \left(160 \gamma ^6-176 \gamma ^5-872 \gamma ^4-188 \gamma ^3+736 \gamma ^2+391 \gamma -21\right) \nonumber \\
&\qquad + 240 (\gamma -1) \left(\gamma ^2-1\right)^{3/2} \left(4 \gamma ^4-12 \gamma ^2-3\right) \operatorname{ArcCosh}(\gamma)\bigg) \nonumber \\
&\qquad + \nu\bigg(4 (\gamma -1) \left(400 \gamma ^7-896 \gamma ^6+284 \gamma ^5+4942 \gamma ^4+2347 \gamma ^3-1671 \gamma ^2-1651 \gamma -755\right) \nonumber \\
&\qquad - 240 \sqrt{\gamma ^2-1} \left(8 \gamma ^7-12 \gamma ^6-12 \gamma ^5+58 \gamma ^4+16 \gamma ^3-12 \gamma ^2-12 \gamma -9\right) \operatorname{ArcCosh}(\gamma)\bigg) \nonumber \\
&\qquad + \frac{\gamma -1}{\Gamma +1}\bigg(40 \left(28 \gamma ^7-240 \gamma ^6-132 \gamma ^5+480 \gamma ^4+183 \gamma ^3-240 \gamma ^2-88 \gamma +15\right) \nonumber \\
&\qquad - 480 \left(\gamma ^2-1\right)^{3/2} \left(4 \gamma ^4-12 \gamma ^2-3\right) \operatorname{ArcCosh}(\gamma) \nonumber \\
&\qquad + \delta\Big(40 (\gamma -1) \left(28 \gamma ^6-116 \gamma ^5-248 \gamma ^4+40 \gamma ^3+223 \gamma ^2+79 \gamma -9\right) \nonumber \\
&\qquad - 480 \left(\gamma ^2-1\right)^{3/2} \left(4 \gamma ^4-12 \gamma ^2-3\right) \operatorname{ArcCosh}(\gamma)\Big) \nonumber \\
&\qquad + \nu\Big(-10 (\gamma -1) \left(336 \gamma ^6-1456 \gamma ^5-3124 \gamma ^4+110 \gamma ^3+2459 \gamma ^2+1364 \gamma +251\right) \nonumber \\
&\qquad + 480 \left(\gamma ^2-1\right)^{3/2} \left(12 \gamma ^4-10 \gamma ^3-36 \gamma ^2-9 \gamma -9\right) \operatorname{ArcCosh}(\gamma) \nonumber \\
&\qquad + \delta\big(-10 (\gamma -1)^2 \left(112 \gamma ^5-800 \gamma ^4-2156 \gamma ^3-1814 \gamma ^2-893 \gamma -299\right) \nonumber \\
&\qquad + 480 (2 \gamma +1) \left(\gamma ^2-1\right)^{3/2} \left(2 \gamma ^3-6 \gamma ^2-3 \gamma -3\right) \operatorname{ArcCosh}(\gamma)\big)\Big)\bigg) \nonumber \\
&\qquad - \frac{120 (\gamma -1)^2 \left(4 \gamma ^3-4 \gamma -1\right) \left(4 \gamma ^3-4 \gamma +1\right) (\delta +1)}{(\Gamma +1)^2}\Bigg], \\
c^{(3)}_{(11),Lp} &= \frac{\pi }{16 \left(\gamma ^2-1\right)^3} \Bigg[-9 + 420 \gamma + 156 \gamma ^2 - 1484 \gamma ^3 - 365 \gamma ^4 + 1120 \gamma ^5 + 210 \gamma ^6 \nonumber \\
&\qquad + (\gamma -1) \left(210 \gamma ^5+1330 \gamma ^4+5 \gamma ^3-1071 \gamma ^2-51 \gamma +105\right) \delta \nonumber \\
&\qquad - 2 (\gamma -1) \left(210 \gamma ^5+265 \gamma ^4-252 \gamma ^3-222 \gamma ^2+42 \gamma +21\right) \nu \nonumber \\
&\qquad - \frac{32 (\gamma -1)}{\Gamma +1}\bigg(-2 \gamma  \left(35 \gamma ^4-40 \gamma ^2+9\right) + \left(150 \gamma ^5-65 \gamma ^4-171 \gamma ^3+57 \gamma ^2+39 \gamma -6\right) \nu \nonumber \\
&\qquad + \gamma  \delta  \left((\gamma -1) (2 \gamma +1) \left(5 \gamma ^2-3\right) \nu -2 \left(35 \gamma ^4-40 \gamma ^2+9\right)\right)\bigg)\Bigg], \\
c^{(3)}_{(12),Lp} &= \frac{\pi }{4 \left(\gamma ^2-1\right)^3} \Bigg[-(\gamma -1) \left(250 \gamma ^4+5 \gamma ^3-222 \gamma ^2-3 \gamma +24\right) \delta \nonumber \\
&\qquad + \gamma  \left(310 \gamma ^4-5 \gamma ^3+2 (\gamma -1) \left(5 \gamma ^2-3\right) \left(10 \gamma ^2-3 \gamma -5\right) \nu -413 \gamma ^2+3 \gamma +117\right) \nonumber \\
&\qquad + \frac{8 (\gamma -1)}{\Gamma +1}\bigg(3 \left(10 \gamma ^4-9 \gamma ^2+1\right) + 3 \left(10 \gamma ^4-9 \gamma ^2+1\right) \delta \nonumber \\
&\qquad + \nu  \left(130 \gamma ^5-55 \gamma ^4-149 \gamma ^3-(\gamma -1) (2 \gamma +1) \left(5 \gamma ^2-3\right) \gamma  \delta +51 \gamma ^2+33 \gamma -6\right)\bigg)\Bigg], \\
c^{(3)}_{\{12\},Lb} &= \frac{2}{3 \left(\gamma ^2-1\right)^4} \Bigg[-9 \left(\gamma ^2-1\right) \left(48 \gamma ^5-24 \gamma ^4-68 \gamma ^3+28 \gamma ^2+17 \gamma -3\right) \nonumber \\
&\qquad + \nu\bigg(-4 \left(\gamma ^2-1\right) \left(80 \gamma ^6-356 \gamma ^5-492 \gamma ^4+200 \gamma ^3+219 \gamma ^2-285 \gamma +175\right) \nonumber \\
&\qquad + 12 \sqrt{\gamma ^2-1} \left(32 \gamma ^7-32 \gamma ^6-208 \gamma ^5+80 \gamma ^3-108 \gamma ^2+96 \gamma -13\right) \operatorname{ArcCosh}(\gamma)\bigg) \nonumber \\
&\qquad + \frac{2 \left(\gamma ^2-1\right)}{\Gamma +1}\bigg(3 \left(64 \gamma ^5-36 \gamma ^4-92 \gamma ^3+41 \gamma ^2+23 \gamma -4\right) \nonumber \\
&\qquad + \nu\Big((\gamma -1) \left(112 \gamma ^5+256 \gamma ^4-668 \gamma ^3+94 \gamma ^2+523 \gamma -347\right) \nonumber \\
&\qquad - 48 (\gamma -1) (2 \gamma -1) \sqrt{\gamma ^2-1} \left(2 \gamma ^3+6 \gamma ^2-3 \gamma +3\right) \operatorname{ArcCosh}(\gamma)\Big)\bigg)\Bigg], \\
c^{(3)}_{[12],Lb} &= -\frac{2 (2 \gamma +1) \delta }{(\gamma +1) \left(\gamma ^2-1\right)^2} \Bigg[1 - 6 \gamma + 8 \gamma ^3 \nonumber \\
&\qquad + \frac{2 \left((\gamma -1) \left(8 \gamma ^3-4 \gamma ^2-4 \gamma +1\right) \nu -\gamma  (2 \gamma -1)\right)}{\Gamma +1}\Bigg].
\end{align}

In addition, we also provide the \emph{covariant} polar scattering angle coefficients. Note that we use a different basis of four vectors, namely $u_1$ and $u_2$ instead of $\hat{P}$ and $p_{\rm rel}$, which simplifies the expressions in the covariant SSC. 
\begin{equation}
    \begin{split}
        \frac{\theta}{\Gamma} &= \sum_{i=1}^{3} \left( \frac{G M}{|b|} \right)^i \Big[ \Big\{ \frac{a_{1,\mu}}{|b|} \left( c^{(i)}_{1,b} \hat{b}^\mu + c^{(i)}_{1,u} (u_2 - u_1)^\mu \right)
            \\ & \qquad + \frac{a_1^\mu a_1^\nu}{|b|^2} \hat{L}_\mu \left( c^{(i)}_{(11),Lb} \hat{b}_\nu + c^{(i)}_{(11),Lu} (u_2 - u_1)_\nu \right)
            + c^{(i)}_{(12),Lu} \frac{a_1^\mu a_2^\nu}{|b|^2} \hat{L}_\mu (u_2 - u_1)_\nu \Big\} + \left( 1 \leftrightarrow 2 \right) \\
        & \qquad + \frac{\hat{L}_\mu \hat{b}_\nu}{|b|^2} \left( c^{(i)}_{\{12\},Lb} \left( a_1^\mu a_2^\nu + a_2^\mu a_1^\nu \right) + c^{(i)}_{[12],Lb} \left( a_1^\mu a_2^\nu - a_2^\mu a_1^\nu \right) \right) \Big]
        \label{eq:theta-expansion}
    \end{split}
\end{equation}
Where appropriate, we use a self-force split of the coefficients, as they are polynomial in the symmetric mass ratio $\nu$:
\begin{equation}
    c^{(i)}_{\bullet} = \sum_{k \geq 0} \nu^k c^{(i,k)}_{\bullet}
    \label{eq:sf-split}
\end{equation}

% ---------------- 1PM ----------------
\begin{align}
c^{(1)}_{1,b} &= -\frac{4 \gamma }{\sqrt{\gamma ^2-1}}, \qquad
c^{(1)}_{1,u} = 0, \qquad
c^{(1)}_{(11),Lb} = -\frac{4 \left(2 \gamma ^2-1\right)}{\gamma ^2-1}, \qquad
c^{(1)}_{(11),Lu} = 0, \\
c^{(1)}_{(12),Lu} &= 0, \qquad
c^{(1)}_{\{12\},Lb} = -\frac{4 \left(2 \gamma ^2-1\right)}{\gamma ^2-1}, \qquad
c^{(1)}_{[12],Lb} = 0.
\end{align}

% ---------------- 2PM ----------------
\begin{align}
c^{(2)}_{1,b} &= -\frac{\pi  \gamma  \left(5 \gamma ^2-3\right)}{8 \left(\gamma ^2-1\right)^{3/2}}\left[7 + \delta\right], \\
c^{(2)}_{1,u} &= -\frac{1}{\left(\gamma ^2-1\right)^2}\left[-1 - 4 \gamma + 4 \gamma ^2 + 8 \gamma ^3 + \left(8 \gamma ^3-4 \gamma ^2-4 \gamma +1\right) \delta\right], \\
c^{(2)}_{(11),Lb} &= -\frac{3 \pi }{32 \left(\gamma ^2-1\right)^2}\left[23 - 162 \gamma ^2 + 155 \gamma ^4 + 7 \left(\gamma ^2-1\right) \left(5 \gamma ^2-1\right) \delta\right], \\
c^{(2)}_{(11),Lu} &= -\frac{3}{\left(\gamma ^2-1\right)^{5/2}}\left[(\gamma +1) \left(8 \gamma ^3-4 \gamma ^2-4 \gamma +1\right) + (\gamma -1) \left(8 \gamma ^3+4 \gamma ^2-4 \gamma -1\right) \delta\right], \\
c^{(2)}_{(12),Lu} &= \frac{3}{\left(\gamma ^2-1\right)^{5/2}}\left[-(\gamma +1) \left(8 \gamma ^3-4 \gamma ^2-4 \gamma +1\right) + (\gamma -1) \left(8 \gamma ^3+4 \gamma ^2-4 \gamma -1\right) \delta\right], \\
c^{(2)}_{\{12\},Lb} &= -\frac{3 \pi  \left(20 \gamma ^4-21 \gamma ^2+3\right)}{4 \left(\gamma ^2-1\right)^2}, \qquad
c^{(2)}_{[12],Lb} = 0.
\end{align}

% ---------------- 3PM ----------------
\begin{align}
c^{(3,0)}_{1,b} &= -\frac{4 \gamma  \left(8 \gamma ^4-12 \gamma ^2+3\right)}{\left(\gamma ^2-1\right)^{5/2}}, \\
c^{(3,1)}_{1,b} &= \frac{8}{3 (\gamma +1) \left(\gamma ^2-1\right)^2} \Bigg[-\sqrt{\gamma ^2-1} \left(10 \gamma ^5-14 \gamma ^4-92 \gamma ^3-56 \gamma ^2-11 \gamma -20\right) \nonumber \\
&\qquad + 6 \gamma  (\gamma +1) \left(\gamma ^2-6\right) \left(2 \gamma ^2+1\right) \operatorname{ArcCosh}(\gamma)\Bigg], \\
c^{(3,0)}_{1,u} &= -\frac{\pi }{4 \left(\gamma ^2-1\right)^3} \Bigg[3 + 18 \gamma - 27 \gamma ^2 - 80 \gamma ^3 + 30 \gamma ^4 + 70 \gamma ^5 
 + \left(70 \gamma ^5-30 \gamma ^4-80 \gamma ^3+27 \gamma ^2+18 \gamma -3\right) \delta\Bigg], \\
c^{(3,1)}_{1,u} &= \frac{\pi  \gamma  (2 \gamma +1) \left(5 \gamma ^2-3\right)}{2 (\gamma +1) \left(\gamma ^2-1\right)^2}, \\
c^{(3,0)}_{(11),Lb} &= -\frac{2}{\left(\gamma ^2-1\right)^3} \Bigg[-5 + 71 \gamma ^2 - 148 \gamma ^4 + 80 \gamma ^6
 + \left(\gamma ^2-1\right) \left(4 \gamma ^2-2 \gamma -1\right) \left(4 \gamma ^2+2 \gamma -1\right) \delta\Bigg], \\
c^{(3,1)}_{(11),Lb} &= \frac{4}{15 \left(\gamma ^2-1\right)^4} \Bigg[-\left(\gamma ^2-1\right) \left(400 \gamma ^7-1200 \gamma ^6-4888 \gamma ^5+2220 \gamma ^4+706 \gamma ^3-1065 \gamma ^2+752 \gamma +75\right) \nonumber \\
&\qquad + 120 \gamma ^2 \sqrt{\gamma ^2-1} \left(4 \gamma ^6-36 \gamma ^4+\gamma ^2+6\right) \operatorname{ArcCosh}(\gamma)\Bigg], \\
c^{(3,0)}_{(11),Lu} &= -\frac{\pi }{16 \left(\gamma ^2-1\right)^{7/2}} \Bigg[(\gamma +1) \left(1330 \gamma ^5-850 \gamma ^4-1275 \gamma ^3+639 \gamma ^2+237 \gamma -57\right) \nonumber \\
&\qquad + (\gamma -1) \left(1330 \gamma ^5+850 \gamma ^4-1275 \gamma ^3-639 \gamma ^2+237 \gamma +57\right) \delta\Bigg], \\
c^{(3,1)}_{(11),Lu} &= \frac{\pi  \left(370 \gamma ^4-185 \gamma ^3-243 \gamma ^2+69 \gamma +21\right)}{8 \left(\gamma ^2-1\right)^{5/2}}, \\
c^{(3,0)}_{(12),Lu} &= \frac{\pi }{4 \left(\gamma ^2-1\right)^{7/2}} \Bigg[-(\gamma +1) \left(280 \gamma ^5-130 \gamma ^4-315 \gamma ^3+114 \gamma ^2+69 \gamma -12\right) \nonumber \\
&\qquad + (\gamma -1) \left(280 \gamma ^5+130 \gamma ^4-315 \gamma ^3-114 \gamma ^2+69 \gamma +12\right) \delta\Bigg], \\
c^{(3,1)}_{(12),Lu} &= -\frac{\pi  \gamma  (2 \gamma -1) \left(5 \gamma ^2-3\right)}{2 \left(\gamma ^2-1\right)^{5/2}}, \\
c^{(3,0)}_{\{12\},Lb} &= -\frac{2 \left(80 \gamma ^6-148 \gamma ^4+71 \gamma ^2-5\right)}{\left(\gamma ^2-1\right)^3}, \\
c^{(3,1)}_{\{12\},Lb} &= \frac{4}{3 \left(\gamma ^2-1\right)^4} \Bigg[-\left(\gamma ^2-1\right) \left(80 \gamma ^7-240 \gamma ^6-1120 \gamma ^5+444 \gamma ^4-2 \gamma ^3-213 \gamma ^2+118 \gamma +15\right) \nonumber \\
&\qquad + 6 \sqrt{\gamma ^2-1} \left(16 \gamma ^8-160 \gamma ^6-20 \gamma ^4+12 \gamma ^2-1\right) \operatorname{ArcCosh}(\gamma)\Bigg], \\
c^{(3)}_{[12],Lb} &= -\frac{2 \left(4 \gamma ^2-2 \gamma -1\right) \left(4 \gamma ^2+2 \gamma -1\right) \delta }{\left(\gamma ^2-1\right)^2}.
\end{align}

\end{document}